\documentstyle[12pt,epsfig,fleqn]{article}
\newcommand{\be}{\begin{eqnarray}} 
\newcommand{\ee}{\end{eqnarray}} 
\newcommand{\nee}{\nonumber\end{eqnarray}} 
\newcommand{\nn}{\nonumber\\} 
\newcommand\sfrac[2]{{\textstyle \frac{#1}{#2}}}
\newcommand{\dgs}{d^{\gamma}{\!\scriptstyle (}s{\scriptstyle )}\,} 
\newcommand{\dzs}{d^{Z}{\!\scriptstyle (}s{\scriptstyle )}\,} 
\newcommand{\dgz}{d^{\gamma,Z}{\!\scriptstyle (}s{\scriptstyle )}\,} 
\newcommand{\mIm}{\,\mbox{\small $\Im$m\,}}

\newcommand{\plmin}
  {\mbox{
   \mbox{\raisebox{-0.1ex}{${\scriptscriptstyle (}\!$}} 
   \mbox{\raisebox{ 0.2ex}{$\!{\displaystyle \pm}\!$}}    
   \mbox{\raisebox{-0.1ex}{$\!{\scriptscriptstyle )}$}}
  }} 
\begin{document}
\begin{titlepage}
\begin{flushright} 
  hep-ph/9803426 \\ 
  HEPHY--PUB 686 \\ 
  UWThPh--1998--8 
\end{flushright}
\vfill
\begin{center}
{\Large\bf CP Violation in $e^{+}e^{-} \to t\bar{t}$: 
       Energy Asymmetries and Optimal Observables of $b$ 
       and $\bar{b}$ Quarks }
\\ 
\vspace{2cm} 
{\large A. Bartl} \\ 
{\em Institut f\"ur Theoretische Physik, Universit\"at Wien, \\ 
 A-1090 Vienna, Austria} \\ 
\vspace{1cm} 
{\large E. Christova} \\ 
{\em Institute of Nuclear Research and Nuclear Energy, \\ 
 Boul. Tzarigradsko Chaussee 72, Sofia 1784, Bulgaria} \\ 
\vspace{1cm} 
{\large T. Gajdosik, W. Majerotto} \\ 
{\em Institut f\"ur Hochenergiephysik der 
 \"Osterreichischen Akademie der Wissenschaften, \\ 
 A-1050 Vienna, Austria} 
\end{center} 
\vfill

\begin{abstract} %
The energy spectra of the $b$ and $\bar{b}$ quarks in 
$e^{+}e^{-} \to t \bar{t}$ with $t \to b W$ 
($\bar{t} \to \bar{b} W$) are calculated assuming CP violation 
in the production process. Expressions for the corresponding 
CP violating asymmetries and optimal observables that determine 
the imaginary parts of the electric and weak dipole moment 
form factors of the top quark are derived. It is shown that 
polarization of the initial lepton beams is essential to disentangle 
$\mIm \dgs$ and $\mIm \dzs$. Numerical results are given 
in the Minimal Supersymmetric Standard Model with complex phases. 
\end{abstract}
\vfill
\end{titlepage}
\newpage
\setcounter{footnote}{0}
\setcounter{page}{1}

There has been considerable interest in testing physics 
beyond the Standard Model (SM) through possible observation of 
CP violation in processes with top 
quarks~\cite{{CP-violation},{energy-asymmetries},{angle-paper}}. 
The reason is that the SM effects are strongly suppressed, 
while there are many models in which CP violation arises 
naturally at one--loop level. The most widely discussed examples 
are supersymmetric models~\cite{susy} and models 
with more than one Higgs doublet~\cite{higgs}.
In~\cite{angle-paper} we have studied CP violating angular 
asymmetries of the $b$ and $\bar{b}$ quarks in the processes:
\be
e^{+} + e^{-} & \to & t + \bar{t} \to b + \bar{t} W^{+}
\; ,
\label{1}
\\
e^{+} + e^{-} & \to & t + \bar{t} \to \bar{b} + t W^{-}
\; .
\label{2}
\ee
In this article we shall discuss the possibility of measuring 
CP non conservation through the energy distributions of 
$b$ and $\bar{b}$ quarks from the top quark decays. 
We also consider the case of longitudinally polarized 
initial beams.

Various CP violating energy asymmetries of the final leptons from 
$t$ and $\bar{t}$ decays have been considered 
in~\cite{energy-asymmetries}. We show that measurements of the 
energies of $b$ and $\bar{b}$ quarks 
provide another possibility of obtaining information about CP 
violation in the $t\bar{t}$ production process. 
CP non conservation in the $t\bar{t}$ production process is induced 
by the electric $\dgs$ and weak $\dzs$ dipole moment form factors in 
the $\gamma t\bar{t}$ and $Z t\bar{t}$ vertices 
\be
  ({\mathcal V}_{\gamma})^{\mu}
&=&
  \sfrac{2}{3} \gamma^{\mu} 
- i ( {\mathcal P}^{\mu} / m_{t} ) \dgs \gamma^{5}
\; , 
\label{photonvertex} 
\\
  ({\mathcal V}_{Z})^{\mu}
&=&
  \gamma^{\mu} ( g_{V} + g_{A} \gamma^{5} )
- i ( {\mathcal P}^{\mu} / m_{t} ) \dzs \gamma^{5}
\; , 
\label{Zvertex} 
\ee
where $g_{V} = (1/2) - (4/3) \sin^{2}\Theta_{W}$, 
$g_{A} = - (1/2)$ are the SM couplings of $Z$ to $t\bar{t}$, 
\mbox{${\cal P}_{\mu} = p_{t\,\mu} - p_{\bar{t}\,\mu}$}, and 
$\Theta_{W}$ is the Weinberg angle.

General expressions for determining the real and imaginary parts of 
$\dgs$ and $\dzs$ separately have been obtained in \cite{angle-paper}. 
We continue this study 
considering the energy distributions of $b$ and $\bar{b}$ that provide 
additional measurements of \mbox{$\mIm \dgs$} and \mbox{$\mIm \dzs$}. 
We get analytic expressions for the energy distributions of
$b$ and $\bar{b}$ and define model independent CP violating 
observables, in particular optimal observables. We give numerical 
results in the Minimal Supersymmetric Standard Model (MSSM) with 
complex parameters, using the results for $\dgz$ of ref.\cite{dipole}.

The energy distributions of the $b$ and $\bar{b}$ quarks can be 
derived from the differential cross section for 
(\ref{1}) and (\ref{2}) obtained in~\cite{angle-paper}. 
Choosing a frame where the  $z$--axis points into the direction 
of the top quark one obtains 
\be 
\frac{d\, \sigma^{b(\bar{b})}_{\lambda\lambda'}}
     {d\, x_{b(\bar{b})}}
&=&
\frac{\pi \alpha_{em}^{2}}{s} 
\frac{m_{t}^{2}}{m_{t}^{2} - m_{W}^{2}} 
\left(
    c_{0}^{b(\bar{b})}
  + c_{1}^{b(\bar{b})} x_{b(\bar{b})}
\right)
\; ,
\label{energy} 
\ee 
where $\lambda$ and $\lambda'$ denote the degree of longitudinal 
beam polarizations of $e^{-}$ and $e^{+}$, respectively, and 
\label{ci}
\be 
  c_{0}^{b(\bar{b})}
\label{c0}
&=&
  N_{tot} + 4 \alpha_{b} ( G_{3} \plmin 2 \mIm H_{1} )
\; ,
\nn
  c_{1}^{b(\bar{b})}
\label{c1}
&=&
- 4 \alpha_{b} 
  \frac{2 m_{t}^{2}}{m_{t}^{2} - m_{W}^{2}} 
  ( G_{3} \plmin 2 \mIm H_{1} )
\; . 
\ee
We have used the conventional dimensionless energy variables 
$x_{b(\bar{b})} = \frac{2 E_{b(\bar{b})}}{\sqrt{s}}$ and the 
notation 
\be
  \alpha_{b} 
&=&
  \frac{m_{t}^{2} - 2 m_{W}^{2}}
       {m_{t}^{2} + 2 m_{W}^{2}}
\; ,
\\
N_{tot} &=& ( 3 + \beta^{2} ) F_{1} + 3 ( 1 - \beta^{2} ) F_{2} 
\; .
\ee
The CP violating contribution is determined by the function $H_{1}$
\be
  H_{1} 
= ( 1 - \lambda \lambda' ) H_{1}^{0} 
+ ( \lambda - \lambda' ) D_{1}^{0}
\; ,
\label{H} 
\ee
where $H_{1}^{0}$ and $D_{1}^{0}$ are two independent linear 
combinations of the dipole moment form factors $\dgs$ and $\dzs$, 
\label{H0}
\be 
H_{1}^{0} 
&=& ( \sfrac{2}{3} - c_{V} g_{V} h_{Z} ) \dgs
  - ( \sfrac{2}{3} c_{V} h_{Z} 
    - ( c_{V}^{2} + c_{A}^{2} ) g_{V} h_{Z}^{2} ) \dzs
\; ,
\nn
D_{1}^{0}
&=& 
  - c_{A} g_{V} h_{Z} \; \dgs
  - ( \sfrac{2}{3} c_{A} h_{Z} 
    - 2 c_{V} c_{A} g_{V} h_{Z}^{2} ) \dzs
\; .
\ee
The SM contribution is determined by the functions~\cite{Draganov} 
\be 
  F_{i} 
&=& ( 1 - \lambda \lambda' ) F_{i}^{0} 
  + ( \lambda - \lambda' ) G_{i}^{0}
\qquad i=1,2,3
\; ,
\label{F} 
\nn
  G_{i} 
&=& ( 1 - \lambda \lambda' ) G_{i}^{0} 
  + ( \lambda - \lambda' ) F_{i}^{0}
\; , 
\label{G} 
\ee 
where 
\label{FG0}
\be
F_{1,2}^{0} 
&=& \sfrac{4}{9} - \sfrac{4}{3} c_{V} g_{V} h_{Z} 
+ ( c_{V}^{2} + c_{A}^{2} ) ( g_{V}^{2} \pm g_{A}^{2} ) h_{Z}^{2}
\; ,
\nn
F_{3}^{0} 
&=& 
- \sfrac{4}{3} c_{A} g_{A} h_{Z} 
+ 4 c_{V} c_{A} g_{V} g_{A} h_{Z}^{2}
\; ,
\nn
G_{1,2}^{0}
&=& 
- \sfrac{4}{3} c_{A} g_{V} h_{Z} 
+ 2 c_{V} c_{A} ( g_{V}^{2} \pm g_{A}^{2} ) h_{Z}^{2}
\; ,
\nn
G_{3}^{0}
&=& 
- \sfrac{4}{3} c_{V} g_{A} h_{Z} 
+ 2 ( c_{V}^{2} + c_{A}^{2} ) g_{V} g_{A} h_{Z}^{2}
\; .
\ee
The quantities 
$c_{V} = - (1/2) + 2 \sin^{2}\Theta_{W}$, 
$c_{A} = (1/2)$ are the SM couplings of $Z$ to the electron and 
$h_{Z} = [ s / ( s - m_{Z}^{2} ) ] / \sin^{2} 2 \Theta_{W}$.

The energy distributions of $b$ and $\bar{b}$ are linear 
functions of $x_{b(\bar{b})}$, eq.(\ref{energy}). 
If CP invariance holds the energy distributions of $b$ and $\bar{b}$
are equal. The sensitivity to CP violation is determined by two 
factors: $\alpha_{b}$ measuring the sensitivity of the $b$ quarks to 
the top quark polarization, and $\mIm \dgz$ entering the top quark 
polarization in the production plane. 

In general CP invariance implies 
\be
  \frac{d\sigma^{b}_{\lambda ,\lambda'}}{d\,x_{b}}
= \frac{d\sigma^{\bar{b}}_{-\lambda' ,-\lambda}}{d\,x_{\bar{b}}}
\; .
\ee
The corresponding integrated energy observable 
${\mathcal A}^{E}_{\lambda\lambda'}$ indicating 
CP violation is the difference between the asymmetries in the 
energy distributions 
\be 
{\mathcal A}^{E}_{\lambda\lambda'}
= R_{\lambda\lambda'}^{b}
- R_{-\lambda'-\lambda}^{\bar{b}}
= - 4 \alpha_{b} \beta \mIm H_{1} / N_{\rm{tot}} 
\; ,
\label{Rall}
\ee
where
\be 
R_{\lambda\lambda'}^{b(\bar{b})}
= \Delta N^{b(\bar{b})}(\lambda,\lambda') 
  / N_{\rm{tot}}^{b(\bar{b})}(\lambda,\lambda') 
\; ,
\ee
\be
\Delta N^{b(\bar{b})}(\lambda,\lambda')
&=& N^{b(\bar{b})}(x > x_{0},\lambda,\lambda') 
- N^{b(\bar{b})}(x < x_{0},\lambda,\lambda') 
\; ,
\label{Rgtlt}
\ee
\be
x_{0} = \frac{m_{t}^{2} - m_{W}^{2}}{2 m_{t}^{2}} 
\; . \quad
\ee
$x_{0}$ is the mean value of the energy, 
$x_{\rm{min}} = x_{0} ( 1 - \beta )$, and 
$x_{\rm{max}} = x_{0} ( 1 + \beta )$. 
$N^{b(\bar{b})}(x>x_{0},\lambda,\lambda')$ is the number of 
$b$($\bar{b}$) quarks with $x>x_{0}$ for the beam polarizations 
$\lambda$ and $\lambda'$, respectively. 

Using polarized electron beams we can define the following 
polarization asymmetry: 
\be
{\mathcal P}^{E}
= R_{P}^{b} - R_{P}^{\bar{b}}
= - 4 \alpha_{b} \beta \mIm D_{1}^{0} / N_{\rm{tot}}^{0} 
\; ,
\label{RPall}
\ee
where
\be
N_{\rm{tot}}^{0} = N_{\rm{tot}} ( \lambda = \lambda' = 0 )
\; ,
\ee
\be
R_{P}^{b(\bar{b})}
&=& 
\frac{( 1 - \lambda \lambda' )}{( \lambda - \lambda' )} 
\times
\frac{\Delta N^{b(\bar{b})}(\lambda,\lambda')
     -\Delta N^{b(\bar{b})}(-\lambda,-\lambda')}
     {N_{\rm{tot}}^{b(\bar{b})}(\lambda,\lambda')
     +N_{\rm{tot}}^{b(\bar{b})}(-\lambda,-\lambda')} 
\; .
\ee
Notice that by measuring the polarization asymmetry eq.(\ref{RPall}) 
one obtains information on $\mIm D_{1}^{0}$, whereas by measuring 
the asymmetry eq.(\ref{Rall}) information on the combination 
eq.(\ref{H}) of $\mIm H_{1}^{0}$ and $\mIm D_{1}^{0}$ can be 
obtained. Only by performing experiments 
with different beam polarizations can one determine both 
$\mIm H_{1}^{0}$ and $\mIm D_{1}^{0}$. 

Optimal observables for extracting information in a statistically 
optimal way were introduced in \cite{{Soni},{Diehl+Nachtmann}}. 
We will apply this method for extracting CP violating quantities in 
the reactions (\ref{1}) and (\ref{2}). We rewrite eq.(\ref{energy}) 
in the form 
\be 
d\, \sigma^{b(\bar{b})}_{\lambda\lambda'}
&=&
( S_{0}^{} \plmin S_{1}^{} \, \mIm H_{1} ) 
\, d\, x_{b(\bar{b})}
\; .
\label{optimal} 
\ee 
The coefficients $S_{0}$ and $S_{1}$ follow 
from eqs.(\ref{energy})--(\ref{FG0}): 
\label{si}
\be 
  S_{0}
\label{s0}
&=&
A \, [ N_{\rm{tot}} + 4 \alpha_{b} G_{3} ( 1 - x/x_{0} ) ]
\; ,
\nn
  S_{1}
\label{s1}
&=&
8 A \, \alpha_{b} G_{3} ( 1 - x/x_{0} )
\; ,
\ee
where $A = \pi \alpha_{em}^{2} /(s \, 2 x_{0})$. 
We choose the optimal observable 
${\mathcal O} = S_{1}^{} / S_{0}^{}$. 
The measurable expectation values are 
\be
\langle {\mathcal O} \rangle^{b(\bar{b})}_{\lambda\lambda'} = 
\frac{\int d\, \sigma^{b(\bar{b})}_{\lambda\lambda'}
               {\mathcal O}}
     {\int d\, \sigma^{b(\bar{b})}_{\lambda\lambda'}}
=
\plmin c_{\lambda\lambda'} \mIm H_{1} 
\; ,
\label{expectation-values} 
\ee
where ``$+$'' is for the $b$ and ``$-$'' for the $\bar{b}$. 
Since $\int d x_{b(\bar{b})} \, S_{1}^{} = 0$, one has 
\be
c_{\lambda\lambda'} = \int d x_{b(\bar{b})} \, S_{0}^{} {\mathcal O}^{2}
\, . 
\label{correlationmatrix} 
\ee
With $a = 4 \alpha_{b} \beta G_{3} / N_{\rm{tot}}$ we have
\be
c_{\lambda\lambda'} 
= 
- \frac{4}{(G_{3})^{2}} 
  \left( 
    1 + \frac{1}{2 a} \log \left| \frac{1 - a}{1 + a} \right| 
  \right)
\; .
\label{covariance}
\ee
Since $\langle {\mathcal O} \rangle^{\bar{b}}_{\lambda\lambda'} = -
\langle {\mathcal O} \rangle^{b}_{\lambda\lambda'}$ we define
\be
\langle \overline{{\mathcal O}} \rangle_{\lambda\lambda'}
= \langle {\mathcal O} \rangle^{b}_{\lambda\lambda'}
- \langle {\mathcal O} \rangle^{\bar{b}}_{\lambda\lambda'}
\; . 
\label{observable} 
\ee
In order to disentangle $\mIm H_{1}^{0}$ and $\mIm D_{1}^{0}$ we use 
the beam polarization. We define 
\be
O_{\lambda\lambda'} 
= c_{\lambda\lambda'}^{-1} 
  \langle \overline{{\mathcal O}} \rangle_{\lambda\lambda'}
\; .
\ee
Measuring $O_{\lambda\lambda'}$ and $O_{-\lambda-\lambda'}$ we get 
two equations for $\mIm H_{1}^{0}$ and $\mIm D_{1}^{0}$ 
and obtain 
\label{HD}
\be
  4 \mIm H_{1}^{0}
&=&
  ( O_{\lambda\lambda'} + O_{-\lambda-\lambda'} ) /
  ( 1 - \lambda \lambda' ) 
\; ,
\label{H10} 
\nn
  4 \mIm D_{1}^{0}
&=&
  ( O_{\lambda\lambda'} - O_{-\lambda-\lambda'} ) /
  ( \lambda - \lambda' )
\; .
\label{D10} 
\ee

So far our expressions for the asymmetries and the optimal observables 
are general and model independent. As an example we present in the 
following results in the MSSM with complex parameters. 
We use the results for the electroweak dipole moment form factors 
of the top quark as obtained in~\cite{dipole}, where we assumed that 
$\mu$, $A_{t}$ and $A_{b}$ are complex. In~\cite{dipole} a complete 
study of the gluino, chargino and neutralino exchanges in the loops 
of \mbox{$\gamma t\bar{t}$} and \mbox{$Z t\bar{t}$} was performed. 
In Fig. 1a,b we show the dependence of 
${\mathcal A}^{E}_{\lambda\lambda'}$ and
${\mathcal P}^{E}$ 
on the c.m.s. energy $\sqrt{s}$ for the beam polarizations 
$\lambda = -\lambda' = \{-0.8,0.,0.8\}$. We take the following 
values for the mass parameters: $M=230$~GeV, $|\mu|=250$~GeV, 
$m_{\tilde{t}_{1}}=150$~GeV, $m_{\tilde{t}_{2}}=400$~GeV,
$m_{\tilde{b}_{1}}=270$~GeV, and $m_{\tilde{b}_{2}}=280$~GeV, and for 
the phases we take $\varphi_{\mu} = \frac{4 \pi}{3}$ and 
$\varphi_{\tilde{t}} = \frac{\pi}{6}$. 
We assume the GUT relation between the gaugino mass parameters: 
$m_{\tilde{g}} = (\alpha_{s}/\alpha_{2}) M \approx 3 M$ and
$M^{\prime} = \sfrac{5}{3} \tan^{2}\Theta_{W} M$.
For the same parameter set we show in Fig. 2 the optimal observable 
$\langle \overline{{\mathcal O}} \rangle_{\lambda\lambda'}$, 
eq.(\ref{observable}), as a function of 
the c.m.s. energy $\sqrt{s}$ for the beam polarizations 
$\lambda = -\lambda' = \{-0.8,0.,0.8\}$.

The spikes are due to thresholds of the intermediate 
particles. As can be seen the dependence on the beam polarization can 
be strong in certain regions of $\sqrt{s}$. As already stated in 
\cite{angle-paper}, to observe an effect of $10^{-3}$ at a future 
$e^{+}e^{-}$ linear collider, an integrated luminosity of 
300~$\rm{fb}^{-1}$ is desired. 

To conclude, we have worked out model independent analytic expressions 
for CP violating energy asymmetries as well as for optimal observables 
of $b$ and $\bar{b}$ quarks in 
$t\bar{t}$ production at an $e^{+}e^{-}$ linear collider. They give 
information on the electric and weak dipole moment form factors of 
the top quark. In order to disentangle $\mIm \dgs$ and $\mIm \dzs$ 
through the energy distribution of the $b$ and $\bar{b}$ quarks
beam polarization is necessary. 

This work has been supported by the 'Fonds zur F\"orderung der 
wissenschaftlichen Forschung' of Austria, project no. P10843--PHY. 
E.C.'s work has been supported by the Bulgarian National Science 
Foundation, Grant Ph--510. E.C. thanks the organizers of the winter 
school '37.~Internationale Universit\"atswochen f\"ur Kern-- und 
Teilchenphysik' for the financial support where this work was 
finished. 


%
\setlength{\unitlength}{1pt}

\begin{figure}
\begin{center}
\begin{picture}(306,309)(0,0)
\put(0,0){\includegraphics{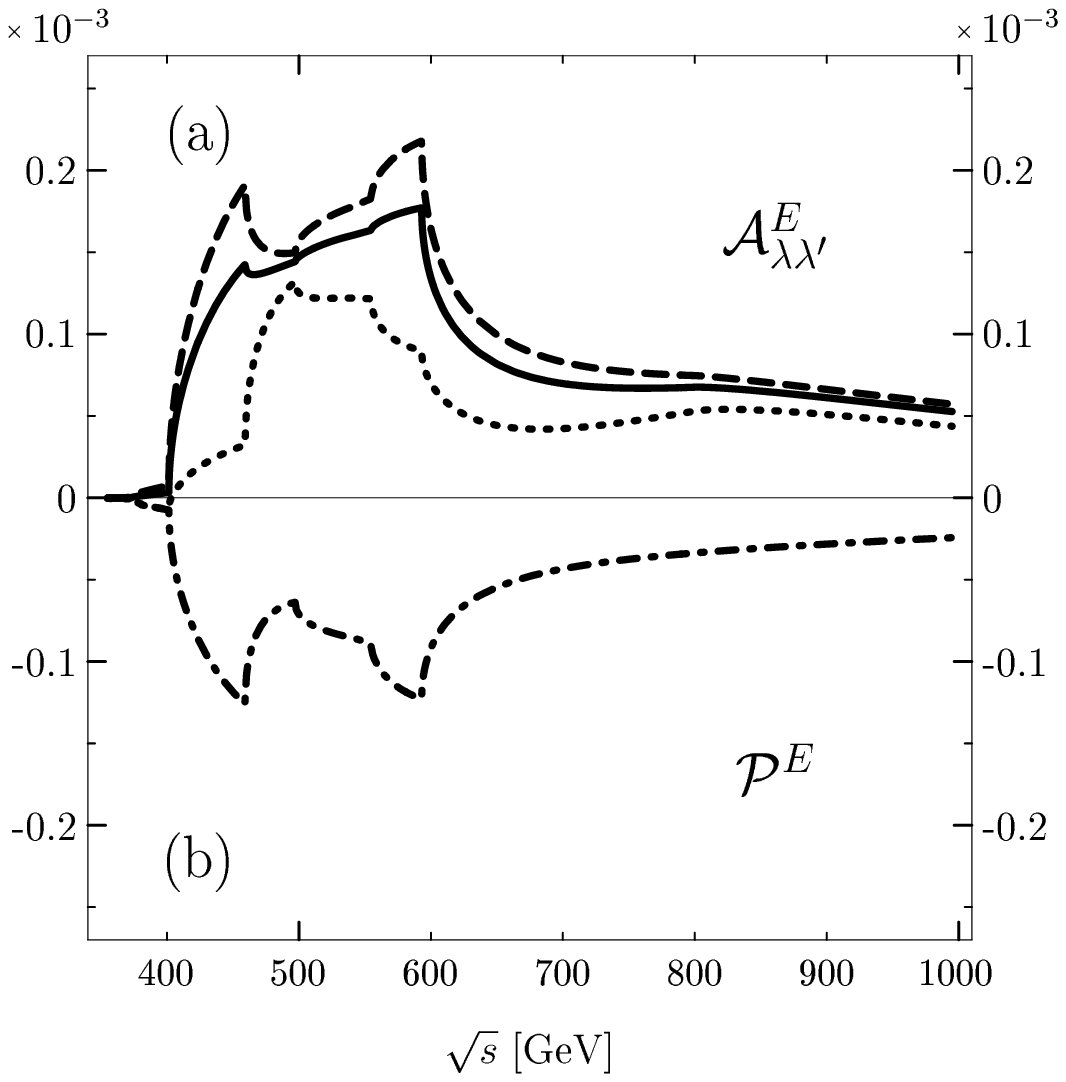}} 
\end{picture}
\end{center}
\caption{(a) the energy asymmetry 
${\mathcal A}^{E}_{\lambda\lambda'}$, eq.(\ref{Rall}), 
as functions of \mbox{$\sqrt{s}$~[GeV]} 
for a longitudinal electron-beam polarization of 
-80\% (dashed line), 
  0\% (full line), and 
 80\% (dotted line) 
and 
(b) the polarization asymmetry ${\mathcal P}^{E}$,
eq.(\ref{RPall}), depending on \mbox{$\sqrt{s}$~[GeV]}
(dashed--dotted line). 
}
\label{fig1}
\end{figure}
\begin{figure}
\begin{center}
\begin{picture}(306,195)(0,0)
\put(0,0){\includegraphics{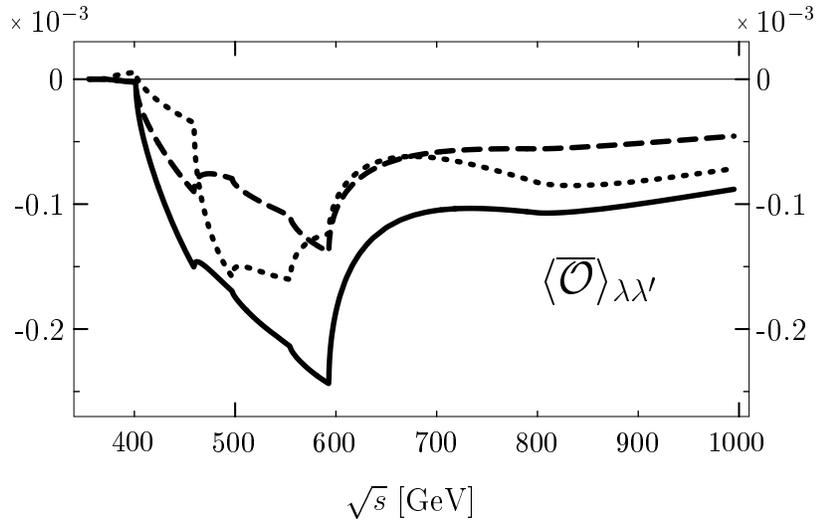}} 
\end{picture}
\end{center}
\caption{The optimal observable 
$\langle \overline{{\mathcal O}} \rangle_{\lambda\lambda'}$, 
eq.(\ref{observable}), as a function of \mbox{$\sqrt{s}$~[GeV]} 
for a longitudinal electron-beam polarization of 
-80\% (dashed line), 
  0\% (full line), and 
 80\% (dotted line).
}
\label{fig2}
\end{figure}
\end{document}